\documentclass[%
 reprint,
superscriptaddress,
 amsmath,amssymb,
 aps,
pre,
]{revtex4-2}

\usepackage{siunitx}
\usepackage{graphicx}
\usepackage{dcolumn}
\usepackage{bm}
\usepackage[export]{adjustbox}

\usepackage{xcolor}

\usepackage{tikz}
\usetikzlibrary{positioning,arrows.meta, shapes.arrows, calc}
\usetikzlibrary{decorations.pathreplacing}
\usetikzlibrary{external}
\definecolor{mplred}{RGB}{255,0,0}
\definecolor{mplgreen}{RGB}{0,127,0}
\definecolor{mplblue}{RGB}{0,0,255}

\usepackage{color}
\definecolor{red}{rgb}{0.8, 0.0, 0.0}

\begin{document}

\preprint{APS/123-QED}
\title{Early Prediction of Creep Failure via Bayesian Inference of Evolving Barriers}
\author{Juan Carlos Verano-Espitia}
\email{juan-carlos.verano-espitia@univ-grenoble-alpes.fr}
\affiliation{Department of Applied Physics, Aalto University, P.O. Box 15600, 00076 Aalto, Espoo, Finland}
\affiliation{Univ. Grenoble Alpes, Univ. Savoie Mont Blanc, CNRS, IRD, Univ. Gustave Eiffel, ISTerre, 38000 Grenoble, France}
\author{Tero Mäkinen}
\affiliation{Department of Applied Physics, Aalto University, P.O. Box 15600, 00076 Aalto, Espoo, Finland}
\author{Mikko J. Alava}
\affiliation{Department of Applied Physics, Aalto University, P.O. Box 15600, 00076 Aalto, Espoo, Finland}
\author{Jérôme Weiss}
\affiliation{Univ. Grenoble Alpes, Univ. Savoie Mont Blanc, CNRS, IRD, Univ. Gustave Eiffel, ISTerre, 38000 Grenoble, France}

\date{\today}

\begin{abstract}
Creep under a sustained load can persist for long times yet culminate in abrupt yielding or rupture, implying a finite lifetime even when the material appears solid. Here, we formulate lifetime prediction as Bayesian inference over an evolving activation-energy landscape. A time-dependent distribution of activation barriers controls deformation: stress lowers barriers, while irreversible rearrangements deplete the weakest sites and reshape the low-barrier tail. Using early-time acoustic emission data, Bayesian inference estimates the evolving barrier statistics in each sample and yields posterior predictive distributions for the time-to-failure. This approach provides online uncertainty-aware lifetime forecasts---already at around 10~\% of the sample lifetime---that link microscopic barrier evolution to macroscopic creep dynamics.
\end{abstract}

\maketitle

Creep---time-dependent deformation under sustained load---often exhibits a long-lived, apparently stable response followed by rapid acceleration to yielding or failure, thereby defining a finite lifetime at fixed loading~\cite{Zhurkov1965,Nechad2005PRL,Brantut2013JSG, Lockwood2024SoftMatter, Divoux2024SoftMatter}. Early prediction of this lifetime remains challenging because the preceding dynamics are slow, history dependent \cite{makinen2023history}, and only weakly informative far from failure~\cite{Nechad2005JMPS,Bell2013GJI}.

A key microscopic concept underlying creep is the distribution of barriers to failure, i.e., the distribution of local distances to instability in activation energy or stress~\cite{lin2014density, lin2014scaling, liu2016driving, ovaska2017excitation, Makinen2023PRM, Korchinski2025PRX, Cao2019PNAS, verano2024heterogeneity}. During primary creep, deformation proceeds through a progressive exploration and depletion of this barrier landscape: the weakest sites fail first, leaving behind a statistically stronger population~\cite{Makinen2023PRM,VeranoEspitia2025PRE}. This generates intrinsic memory effects in the disorder and leads to sample-specific trajectories through the evolving landscape, distinguishing creep from standard Markovian depinning descriptions~\cite{Fisher1998PhysRep,Chauve2000PRB,Ferrero2021ARCMP, makinen2025crack}. While recent work has clarified the role of barrier statistics and aging in transient creep~\cite{makinen2023history, shohat2023logarithmic, Korchinski2025PRX, shohat2026aging, rudyak2025spatiotemporal}, comparatively less attention has been paid to the accelerating regime preceding failure and to the origin and predictability of the associated finite-time singularity~\cite{Nechad2005PRL,SornetteHelmstetter2002PRL}.

Acoustic emission (AE) measurements provide a direct connection to this microscopic picture~\cite{Lockner1993AE,Brantut2013JSG}. In brittle and quasi-brittle materials, AE bursts correspond to individual microcracking or damage events~\cite{Lockner1993AE}, and the statistics of their occurrence have been shown to probe the evolving barrier distribution~\cite{Makinen2023PRM,VeranoEspitia2025PRE}. In particular, the timing of AE events encodes information about the history-dependent state of the system, suggesting that AE activity can be used to infer the evolving landscape and anticipate failure~\cite{VeranoEspitia2025PRE,ZhangZhou2020JGR}.

Here, we formulate creep lifetime prediction as a Bayesian inference problem on this evolving landscape. We treat key descriptors of the barrier distribution as latent variables and infer their posterior distributions from early-time AE observations. This yields posterior predictive distributions for the future creep evolution and the time-to-failure, including uncertainty quantification, and is a tool for online failure prediction. In this sense, the AE event sequence acts as a compact encoding of the non-Markovian memory of the system~\cite{VeranoEspitia2025PRE}. The resulting landscape-inference framework provides a quantitatively testable bridge between microscopic barrier evolution and macroscopic lifetime statistics~\cite{Nechad2005PRL,VeranoEspitia2025PRE}.

From a macroscopic perspective, creep deformation is commonly described by primary, secondary, and tertiary regimes~\cite{Nechad2005JMPS,Brantut2013JSG}, which in many materials collapse onto a master curve under appropriate rescaling and give rise to empirical relations (such as the Monkman--Grant relation~\cite{MonkmanGrant1956}), linking e.g.~the strain-rate minimum to the failure time~\cite{Nechad2005PRL, leocmach2014creep, koivisto2016predicting}. As a baseline, we apply an analogous Bayesian inference approach~\cite{makinen2026growth, Mototake2020SciRep} directly to strain data for each sample. This allows us to assess the predictive gain obtained by incorporating microscopic AE information beyond what is accessible from macroscopic observables alone.

From a statistical-physics viewpoint, creep failure corresponds to a finite-time singularity in the activity (creep rate), driven by damage localization and stress redistribution following local failure events~\cite{Nechad2005PRL,Pradhan2010RMP,SornetteHelmstetter2002PRL, Chaudhuri2025SciPost, castellanos2018creep}. This phenomenon is central to a wide range of systems, including geophysical hazards such as landslides~\cite{intrieri2019forecasting} and slope creep~\cite{Fukuzono1985}, rockfalls~\cite{Amitrano2005GRL}, serac collapses~\cite{Faillettaz2011JGlaciol,Faillettaz2015RevGeophys}, and volcanic eruptions~\cite{Voight1988Nature}. The central difficulty in failure prediction is that early-time data carry information about the failure time but with large uncertainty, whereas accurate predictions typically emerge only close to failure, when intervention is no longer feasible~\cite{Bell2011GRL,Bell2013GJI, Cho2022SoftMatter}.

In this Letter, we show that Bayesian inference based on AE activity enables reliable online prediction of creep failure already in the primary creep stage (as early as at 10~\% of the sample lifetime), significantly earlier than existing methods, while providing fully quantified uncertainties. 
\\

\emph{Methods}---%
Experimental data comes from Ref.~\cite{VeranoEspitia2025PRE}, as the result of 39 tensile creep experiments on sheets of paper (height 100~\si{\milli \meter}, width 50~\si{\milli\meter}). The creep load varies between 220 and 280 \si{\newton}. The collected data includes force and displacement data (used to compute the engineering strain, $\epsilon$), as well as a continuous AE signal.
A catalog of AE events was obtained from the continuous signal using standard thresholding methods, yielding event times $t_j$.

We perform Bayesian inference using Markov chain Monte Carlo (MCMC) methods in the PyMC library~\cite{pymc2023}, specifically the no-U-turn sampler (NUTS). The inference is performed using four independent chains, each with 5000 samples, and a target acceptance rate of 0.9. \\

\begin{figure}[tb!]
    \centering
    \includegraphics[width=1.0\columnwidth]{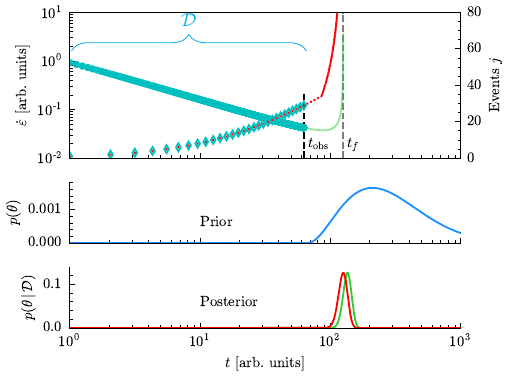}
    \caption{A schematic description of the prediction process, where the top panel shows the continuous strain rate $\dot{\varepsilon}$ (green line) and the discrete accumulated number of events (red dots) as a function of time $t$. The data $\mathcal{D}$ consist of either the strain rate data or the event times read up to a time $t_{\rm obs}$, and based on this data, one tries to predict the failure time $t_{\rm f}$.
    For Bayesian inference, one needs a prior distribution $p(\theta)$, e.g., for the failure time (middle panel).
    By conditioning on the data $\mathcal{D}$, one obtains a posterior distribution $p(\theta \, | \, \mathcal{D})$ (bottom panel), which provides an estimate of the failure time and its associated uncertainty.} 
    \label{fig:schematic}
\end{figure}

\emph{Master curve behavior}---%
For brittle creep, resulting from the accumulation of damage or microcracking events~\cite{brantut2012micromechanics}, the master curve onto which all individual creep curves can be scaled consists of a power-law decrease of the strain rate in primary creep, an inflection point (secondary creep), and a power-law increase in the strain rate towards a finite-time singularity (see top panel of Fig.~\ref{fig:schematic}). This master curve gives a model for the strain rate
\begin{equation} \label{eq:mastercurve}
    \dot{\varepsilon}_{\rm m} = A \left( \frac{t}{t_{f}} \right)^{-p} + B \left( \frac{t_{f} - t}{t_{f}} \right)^{-\gamma}
\end{equation}
where $A$ is a constant prefactor for the primary creep behavior, $t$ the time, $t_{f}$ the failure time, $p$ the primary creep exponent, $B$ a constant prefactor for the tertiary creep, and $\gamma$ is the tertiary creep exponent. To reproduce the known Monkman--Grant-like~\cite{MonkmanGrant1956} relation~\cite{Nechad2005PRL, koivisto2016predicting} between the time of the strain rate minimum and the failure time, $B$ is fixed so that the strain rate minimum occurs at $t=\kappa t_{f}$ for some constant $\kappa$. For our material and geometry, we know that $\kappa \approx 0.83$~\cite{koivisto2016predicting}.

The master curve behavior provides a way to predict the sample's failure from the minimum strain rate. From the primary creep behavior, $A$ and $p$ can be inferred, and if one observes an inflection point in the strain rate, the Monkman--Grant-like relation then allows one to estimate the failure time. In practice, determining the inflection point in a noisy strain-rate signal during an experiment is a formidable task, with significant uncertainty.

Due to this need for uncertainty quantification, Bayesian inference is a natural tool for this task. It is based on the Bayes formula
\begin{equation}\label{eq:bayes}
    p(\theta \, | \, \mathcal{D}) = \frac{p(\mathcal{D} \, | \, \theta) p(\theta)}{\int p(\mathcal{D} \, | \, \theta) p(\theta) \, \mathrm{d}\theta},
\end{equation}
where $\theta$ (here meaning the set $\{ A, p, \kappa, \gamma, t_{f} \}$) denotes the parameters of the model, $\mathcal{D}$ the observed data (for example the observed strain or strain rate data), $p(\theta)$ is the prior distribution of the parameters, $p(\mathcal{D} \, | \, \theta)$ the likelihood, $p(\theta \, | \, \mathcal{D})$ the posterior distribution of the parameters, and the integral is taken over all the possible parameter values. The core idea is to start with estimates of the parameter distributions (priors) and arrive at the parameter distributions conditional on the observed data (posteriors). This idea, applied to creep failure prediction, is illustrated in Fig.~\ref{fig:schematic}.

\begin{figure}[tb]
    \centering
    \includegraphics[width=\columnwidth]{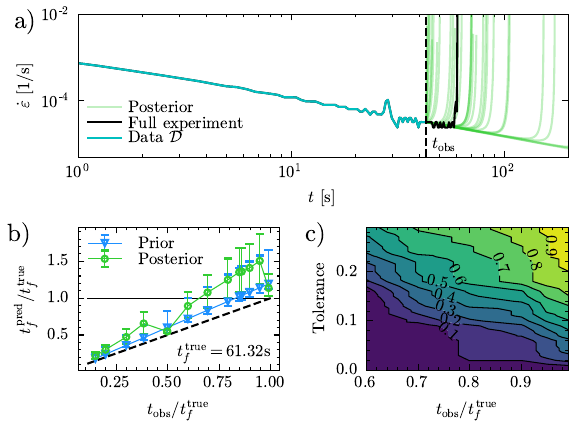}
    \caption{\textbf{a})~Failure time prediction from the strain rate signal for a representative experiment with $t_f^{\rm true}=61.32$~s.
    \textbf{b})~Evolution of the posterior distribution of the failure time (green) as one changes the time $t_{\rm obs}$ (black dashed line) for a representative experiment. The blue symbols show the prior distributions for the failure time, and the error bars represent the 95\% confidence interval.
    \textbf{c})~The proportion of predicted experiments (for 10~\% precision) for a given tolerance and $t_{\rm obs}$.} 
    \label{fig:strain_pred}
\end{figure}

To utilize this framework in creep prediction consists of taking a timeseries of strain data up to time $t_{\rm obs}$, using this as data $\mathcal{D}$ for the Bayesian inference, and seeing what the posterior prediction for the failure time $t_{\rm f}$ is (see Fig.~\ref{fig:strain_pred}a). 
For this, the model of Eq.~\ref{eq:mastercurve} is used, by integrating it up to time $t_{\rm obs}$ [as $\varepsilon_{\rm m}(t_{\rm obs}) = \varepsilon_{\rm m}^0 + \int_{t_0}^{t_{\rm obs}} \dot{\varepsilon}_{\rm m}(t) \,\mathrm{d}t$ where $\varepsilon_{\rm m}^0$ is the strain at the starting point $t_0$]. Gaussian variability (with standard deviation $\sigma_\varepsilon$) around this model is allowed by choosing the likelihood
\begin{equation}
    p\left( \mathcal{D} \, | \, \theta \right)
    =
    \prod_{k=1}^{j_{\rm obs}} \frac{1}{\sqrt{2 \pi \sigma_\varepsilon^2}} \exp\left( - \frac{\left[ \varepsilon_k - \varepsilon_{\rm m}(t_k) \right]^2}{2 \sigma_\varepsilon^2} \right)
\end{equation}
where the index $k$ runs through all the discrete $(t_k, \varepsilon_k)$ datapoints and $j_{\rm obs}$ is the last index satisfying $t_{j_{\rm obs}} \leq t_{\rm obs}$.
It essentially means that the inference aims to minimize the variability around the model curve while still accounting for the observed data.
As the integral in Eq.~\ref{eq:bayes} is not analytically tractable in the general case, sampling from the posterior distribution is done using MCMC methods.
For priors we chose uniform distributions for the exponents, $p \in [\frac{1}{2}, 1]$ and $\gamma \in \left[ \frac{4}{5}, 2 \right]$, and lognormal distribution for the prefactor $A$ (same value for the mean and the standard deviation, $10^{-3}$), common range of values for $p$, $\gamma$ and $A$ in the studied material \cite{koivisto2016predicting, makinen2020scale}. As we know that failure occurs at time $t_{f} > t_{\rm obs}$ and additionally wanted the inference to be set up to catch the earliest signs of tertiary creep onset, we chose a lognormal prior for $t-t_{\rm obs}$, with a mean assuming that we are exactly at the strain rate minimum, and a standard deviation equal to half the mean.

Each sample from the posterior distributions of the parameters produces a full strain rate curve up to time~$t_f$ (Fig.~\ref{fig:strain_pred}a). A more useful metric is the behavior of the prior and posterior distribution (Fig.~\ref{fig:strain_pred}b shows the mean and the 95\% confidence intervals of both for one representative experiment) as a function of $t_{\rm obs}$. The prior mean is always just after $t_{\rm obs}$. Initially, when there is no sign of an inflection point in the strain rate, the posterior mean is much higher than the prior, and the distribution is wider. However, when the inflection point is reached, the posterior assumes a narrower shape and converges to the real $t_{f}$. 
One can also see that a small spike in the strain rate (around 30~s in the experiment of Fig.~\ref{fig:strain_pred}a) can trick the inference and make it predict almost immediate failure ($t_{\rm obs}/t_{\rm f}^{\rm true} = 0.50$ in Fig.~\ref{fig:strain_pred}b), but after the spike the predictions return back to normal.

Looking at this over all the experiments, we can define a precision metric, $\delta t_{f}^{\rm pred} / t_{f}^{\rm true}$ (the narrowness of the posterior distribution, where $\delta$ denotes the standard deviation), and an accuracy metric $\| t_{f}^{\rm true} - \langle t_{f}^{\rm pred} \rangle \| / t_{f}^{\rm true}$ (distance of the posterior mean $\langle t_{f}^{\rm pred} \rangle$ from the true value) and consider a prediction successful if the accuracy is below a given tolerance. 
By varying the tolerance for a constant precision threshold (set as 10~\%), we see~(Fig.~\ref{fig:strain_pred}c) that indeed predictability arises only after the strain-rate minimum. For this material, this means the strain-based method can predict failure only when about 10~\% of the lifetime remains. However, it is useful as a baseline, which will be compared with a self-consistent Bayesian inference method based on the non-Markovian rank-persistence of a catalog of AE event times $t_j$~\cite{VeranoEspitia2025PRE}, from which it is possible to get a creep lifetime prediction already during the primary creep stage.\\

\emph{Acoustic Emission correlations}---%
Instead of the strain data, one can also observe the AE events.
In Ref.~\cite{VeranoEspitia2025PRE} a power-law correlation between the failure time and the time of the $j$th event was found
\begin{equation} \label{eq:correlation}
    t_{f} = e^{C_j} t_j^{\theta_j}
\end{equation}
which held starting from the first $j$s. From the rank-persistence and lognormal nature of the event occurrence time distributions, the relations 
$
    \theta_j = \delta \ln t_{f} / \delta \ln t_j
$
and
$
    C_j = \left\langle \ln t_{f} \right\rangle - \left\langle \ln t_j \right\rangle \theta_j
$
were derived.
These depend on the population averages (denoted $\langle * \rangle$) and standard deviations (defined as $\delta * = \sqrt{\langle ( * - \langle * \rangle )^2 \rangle}$), which one naturally does not know when considering single experiments.

This correlation was explained~\cite{VeranoEspitia2025PRE} by invoking a multiplicative process
\begin{equation} \label{eq:multiplicative}
    t_{j+1} = t_1 \prod_{k=1}^j \left( 1 + r_k \right)
\end{equation}
where $r_k$ is called the multiplier. 
From the observed primary creep evolution of the multiplier $r_k = R_0/k$ 
(where $R_0$ is a constant containing the sample-to-sample variability), one can derive estimates 
for the variables $\theta_j$ and $C_j$ in Eq.~\ref{eq:correlation} (see End Matter for the full derivation). One arrives at functions $\theta_j=\theta_j(N, \delta R_0, \delta \ln t_1)$, (where $N$ roughly corresponds to the total number of events in the experiment) and $C_j=C_j(\langle \ln t_1 \rangle, \langle R_0 \rangle, N, \theta_j)$ which only depend on the distribution of three quantities: $N$, $R_0$, and $t_1$ (see End Matter for details and exact functional forms).\\

\emph{Barrier landscape trajectories and non-Markovian variability}---%
The multiplicative process of Eq.~\ref{eq:multiplicative} links AE event timings to the exploration of an evolving barrier landscape~\cite{VeranoEspitia2025PRE}. Early events activate the weakest sites, while later events probe progressively higher barriers, with $R_0$ controlling the rate of this exploration and capturing intrinsic sample-to-sample variability.

Crucially, this variability resides in the barrier distribution itself, not only in stochastic realizations. Each sample is thus characterized by its own $\{R_0,t_1,N\}$ and follows a distinct trajectory through the landscape, giving rise to the observed correlation (Eq.~\ref{eq:correlation}). As a result, creep dynamics are intrinsically non-Markovian: future evolution depends on the already-depleted portion of the barrier population.

The empirical correlation can be interpreted as a projection of these trajectories onto the $(t_j,t_{f})$ plane, with coefficients $(\theta_j,C_j)$ set by population-level statistics. While these are not directly observable, the AE sequence constrains the admissible parameter space and enables inference of the individual trajectory.
In contrast to strain, which remains insensitive to these differences until late times, the AE signal probes the underlying dynamics and carries predictive information from early stages onward.\\

\emph{Self-consistent Bayesian inference}---%
To apply these ideas to actual prediction, we again set up a Bayesian inference-based prediction scheme.
The key step here is to treat these population statistics $\{R_0,t_1,N\}$ as latent parameters and infer them from the AE timing data.
Taking a logarithm of Eq.~\ref{eq:correlation},
and remembering that the event occurrence times follow a universal lognormal distribution~\cite{VeranoEspitia2025PRE}, this idea of a self-consistent prediction is easily formalized by the Gaussian likelihood for $\ln t_{\rm f}$
\begin{equation}
    p\left( \mathcal{D} \, | \, \theta \right)
    =
    \prod_{j=1}^{j_{\rm obs}} \frac{\exp\left[ - \frac{\left( \langle \ln t_{f} \rangle - C_j - {\theta_j} \ln t_j \right)^2}{2 \sigma_{\ln t_{f}}^2} \right]}{\sqrt{2 \pi \sigma_{\ln t_{f}}^2}}
\end{equation}
where $\mathcal{D}$ now corresponds to the observed $t_j$ for $j \leq j_{\rm obs}$.
This likelihood essentially tries to be self-consistent, i.e.~to minimize the scatter in the predictions for $\ln t_{f}$ (characterized by $\sigma_{\ln t_{f}}$) over all the $j$s up to $j_{\rm obs}$. The difference to the earlier strain-based prediction is that we are not trying to fit the model~(Eq.~\ref{eq:correlation}) as well as possible (with the given $\theta_j$ and $C_j$ functions), but are just trying to have self-consistent failure time predictions.

To set the priors, we assume that the first event is roughly at 1~s (so we set a lognormal prior~\cite{verano2025barrier} with $\langle \ln t_1 \rangle = 0$) and based on previous results~\cite{VeranoEspitia2025PRE} $\langle R_0 \rangle \sim 1$. The sample-to-sample variability is reflected in $t_1$, so for the combined quantity $\delta R_0 / \delta \ln t_1$, we assume $\delta\ln t_1 \sim \delta R_0$ and set a Gaussian prior (mean 1, standard deviation 0.2).
Finally, the most important quantity $N$ has a lognormal prior (mean 250, standard deviation 50). This quantity depends on the AE signal chain and the thresholding made to such a signal. See End Matter for additional discussion on the effects of this prior on the results.\\ 

\begin{figure}[tb]
    \centering
    \includegraphics[width=1.0\columnwidth]{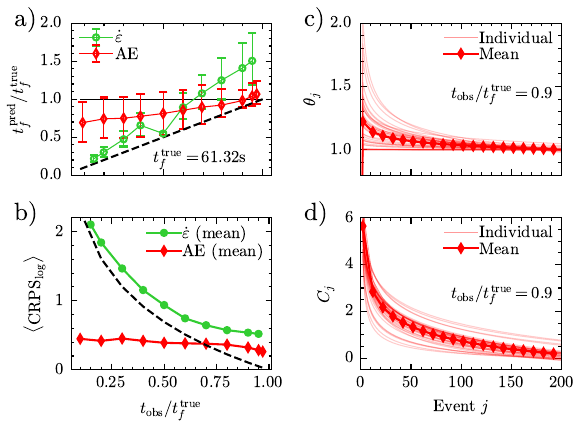}
    \caption{
    \textbf{a})~Comparison of the evolution of the posterior distribution of the failure time as one changes the prediction time $t_{\rm obs}$ (black dashed line) for one single experiment using either the strain rate (green) or AE (red) based method. The error bars represent the 95\% confidence interval.
    \textbf{b})~Mean prediction score $\langle \rm CRPS_{log} \rangle$ over the population of samples from the strain rate (green) or AE (red) based predictions. The black dashed line corresponds to the single-valued $t_f^{\rm{pred}}=t_{\rm obs}$.
    \textbf{c})~Individual trajectories (red lines) for all the experiments and average trajectory (red diamonds) for the exponent~$\theta_j$, inferred at $t_{\rm obs} = 0.9 t_f^{\rm true}$.
    \textbf{d})~Same as panel c but for the prefactor~$C_j$.} 
    \label{fig:ae_pred}
\end{figure}

\emph{Early creep failure prediction}---%
Using this self-consistent Bayesian inference, we can achieve much better early-failure-time predictions (see Fig.~\ref{fig:ae_pred}a for a representative experiment).
The posterior distribution for $t_{\rm f}$ already roughly matches the real failure time at around 10~\% of lifetime, i.e., significantly before the minimum strain rate that occurs (for this material) after 80~\% of lifetime, far surpassing the predictive capabilities based on the strain signal. To quantify the distance between the true failure time and the posterior distribution from the Bayesian inference, we compute the continuous ranked probability score (CRPS) \cite{hersbach2000CRPS} in log-space as
\begin{equation}
    {\rm CRPS}_{\rm log} = 
    \int_0^\infty \frac{\left[ P(t_f^{\rm pred}) - H(\ln t_f^{\rm pred} - \ln t_{f}^{\rm true}) \right]^2}{t_f^{\rm pred}} \mathrm{d} t_f^{\rm pred}
\end{equation}
where $P$ is the cumulative distribution function of the posterior (which we assume to be lognormal~\cite{VeranoEspitia2025PRE} and use the sample mean and standard deviation), and $H$ is the Heaviside step function. 
This metric shows that the self-consistent Bayesian inference presents a better score CRPS${_{\rm log}}$ from much earlier stages (red lines in Fig.~\ref{fig:ae_pred}b) than the strain-based prediction (green lines in Fig.~\ref{fig:ae_pred}b), 
whereas the strain-based prediction attains a reasonable score only after the strain-rate minimum is reached, i.e.~when approaching failure. Notably, the behavior of the CRPS-score for the strain prediction is close to the naive single-valued prediction $t_f^{\rm pred} = t_{\rm obs}$ (dashed line in Fig. \ref{fig:ae_pred}b).

The individual trajectories for each sample can directly be seen by plotting the evolution of inferred $\theta_j$ (Fig.~\ref{fig:ae_pred}c) and $C_j$ (Fig.~\ref{fig:ae_pred}d) for each experiment, based on the posterior means obtained at $t_{\rm obs} = 0.9 t_f^{\rm true}$. Even though they are based on population-level correlations, interpretation using our Bayesian inference scheme yields a unique curve for each experiment.
This illustrates the idea of projection from unique landscape trajectories to unique behaviors in these quantities. The mean of these curves shows a similar pattern to previous results~\cite{VeranoEspitia2025PRE}.
The trajectory evolution is effectively deterministic in $(\theta_j, C_j)$ space, governed by constrained dynamics, while sample-specific disorder enters only via the latent parameters.\\

\emph{Conclusions}---%
We have presented an online Bayesian inference method for predicting creep rupture from microscopic information carried by acoustic emission event timing. 
The predictability arises from intrinsically non-Markovian creep dynamics, in which the statistics of future events depend on both the depletion of weak barriers and the sample-specific disorder ensemble, leading to ranked trajectories in barrier-landscape space. 
By inferring the latent parameters governing these trajectories through a self-consistent Bayesian scheme, early AE activity is converted into a posterior distribution for the failure time with quantified uncertainty. 
AE thus acts not merely as an empirical precursor signal, but as an indirect probe of the evolving barrier landscape governing creep rupture.

The resulting predictions become accurate already at early stages of damage accumulation, well before the predictability implied by the master curve behavior of the strain signal, indicating that the relevant barrier-landscape information is encoded in the earliest events.
The trajectories evolve in a deterministic manner through the constrained dynamics of $\theta_j$ and $C_j$, with sample-specific variability entering only through the underlying latent parameters.

More broadly, the approach provides a route to early failure forecasting in systems governed by collective damage dynamics and finite-time singularities, where macroscopic observables are late and noisy but event-level activity retains microscopic information about the evolving disorder. This trajectory-based inference perspective should be applicable beyond laboratory creep, including geophysical rupture phenomena where precursory event sequences provide a window into the underlying instability.\\

\emph{Acknowledgments}---%
ISTerre is part of Labex OSUG@2020. This work has been supported by the French National Research Agency in the framework of the "Investissements d'Avenir" program (ANR-15-IDEX-02). 
J.C.V.E. thanks funding from the Vilho, Yrjö and Kalle Väisälä Foundation of the Finnish Academy of Science and Letters. 
J.C.V.E., T.M. and M.J.A. acknowledge the support from FinnCERES flagship (grant no.~151830423), Business Finland (grant nos.~211835, 211909, and 211989), and the Research Council of Finland (grant nos.~13359905 and 13361245). 
M.J.A. acknowledges support from the Academy of Finland Center of Excellence program (program nos.~278367 and 317464), as well as the Finnish Cultural Foundation. 
The authors acknowledge the computational resources provided by the Aalto University School of Science “Science-IT” project.


\bibliography{apssamp}


\section*{End Matter}\label{end_matter}

\subsection*{Derivation of the estimates of the population quantities}
We start from the multiplicative process (Eq.~\ref{eq:multiplicative})
and a functional form for the multiplier $r_k = R_0/k$ where $R_0$ is a prefactor accounting for the sample-to-sample variation (having a mean value $\langle R_0 \rangle$ and standard deviation $\delta R_0$). This functional form naturally holds only for primary creep~\cite{VeranoEspitia2025PRE} ($r_k$ decreases rapidly when entering tertiary creep). The sample-to-sample variation could be introduced in other ways (e.g.~varying the power-law exponent of $r_k$ around unity), but the results would be similar and this derivation is in any case an approximation.

By taking the logarithm of Eq.~\ref{eq:multiplicative} and using the expansion $\ln(1+x) \approx x$, one gets
\begin{equation}
    \ln t_{j+1}
    = \ln t_1 + R_0 H_j
\end{equation}
where $H_j = \sum_{k=1}^j k^{-1}$. This has known asymptotics, $H_j \sim \gamma + \ln j$ (where $\gamma \approx 0.577$ is the Euler--Mascheroni constant). With this, and by shifting the indices by one, one gets the form
\begin{equation}
    \ln t_{j} \approx \ln t_1 + R_0 \gamma + R_0 \ln (j-1) .
\end{equation}

Taking the population average over the sample-to-sample random variables ($t_j$, $t_1$, and $R_0$) yields
\begin{equation} \label{eq:mean_tj}
    \left\langle \ln t_{j} \right\rangle = \left\langle \ln t_1 \right\rangle + \gamma \left\langle R_0 \right\rangle + \left\langle R_0 \right\rangle \ln (j-1) .
\end{equation}
Similarly, for the standard deviation (neglecting any covariance terms)
\begin{equation}
    \delta \ln t_{j} \approx \sqrt{\left( \delta \ln t_1 \right)^2 + \left[ \gamma + \ln (j-1) \right]^2 \left( \delta R_0 \right)^2} .
\end{equation}

This enables then the determination of the exponent
\begin{equation}
    \theta_j = \frac{\delta \ln t_{\rm f}}{\delta \ln t_j}
    = \frac{\frac{\delta \ln t_{\rm f}}{\delta \ln t_1}}{\sqrt{1 + \left[ \gamma + \ln (j-1) \right]^2 \left( \frac{\delta R_0}{\delta \ln t_1} \right)^2}}
\end{equation}
which goes to zero as $j \to \infty$. However, we know that $\theta_j$ goes to unity at some finite $j=N$ (close to failure), so we can simply write
\begin{equation}
    \theta_j 
    = \sqrt{\frac{1 + \left[ \gamma + \ln (N-1) \right]^2 \left( \frac{\delta R_0}{\delta \ln t_1} \right)^2}{1 + \left[ \gamma + \ln (j-1) \right]^2 \left( \frac{\delta R_0}{\delta \ln t_1} \right)^2}}
\end{equation}
which only depends on the constants $N$, $\delta R_0$, and $\delta \ln t_1$.

Similarly, we can approximate the mean (log) failure time from Eq.~\ref{eq:mean_tj} by assuming that as $j \to N$, $t_j \to t_{\rm f}$. This enables the computation of the prefactor
\begin{equation}
    \begin{split}
        C_j = &\left( \left\langle \ln t_1 \right\rangle + \gamma \left\langle R_0 \right\rangle \right) (1 - \theta_j) \\
    &+ \left\langle R_0 \right\rangle \left[ \ln (N-1) - \theta_j \ln (j-1) \right]
    \end{split}
\end{equation}
which only depends on $\left\langle \ln t_1 \right\rangle$, $\left\langle R_0 \right\rangle$, and $N$. So by knowing the distributions of $N$, $R_0$, and $t_1$, one can get the full behavior of $\theta_j$ and $C_j$.\\

\begin{figure}[tb!]
    \centering
    \includegraphics[width=1.0\columnwidth]{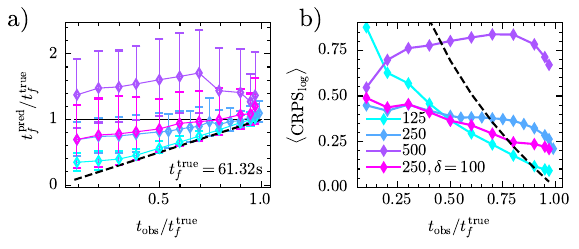}
    \caption{Bayesian inference results (as in Fig.~\ref{fig:ae_pred}) but for different prior distributions of $N$. The colors represent the mean--standard deviation pairs $\{\langle N\rangle, \delta N\}$: cyan \{125, 25\}, blue \{250, 50\}, purple \{500, 100\}, and magenta \{250, 100\}. The black dashed line corresponds to the single-valued prediction $t_f^{\rm{pred}}=t_{\rm obs}$.
    \textbf{a})~The evolution of the prior and posterior distributions of the lifetime as a function of $t_{\rm obs}$ for one single experiment, with the error bars representing the 95\% confidence interval.
    \textbf{b})~The evolution of the mean prediction score $ \langle \rm{CRPS_{log}} \rangle$ over the population of samples. 
    } 
    \label{fig:ae_pred_N}
\end{figure}

\subsection*{Sensitivity of the prediction to the priors}

To assess the sensitivity of the inference to the prior distribution, we systematically vary the prior for the parameter $N$, which we find to be the most influential latent variable in the model. Specifically, for the same lognormal prior we shift the prior mean to twice and to half of its baseline value, while keeping the coefficient of variation (standard deviation divided by the mean) fixed at 0.20.

We observe a clear effect of this prior on the predictions (see Fig.~\ref{fig:ae_pred_N}a). When the prior mean is set significantly above the optimal value, the inferred failure time $t_f^{\rm pred}$ is slightly overestimated (although by less than the factor of two the prior mean was shifted by), leading to persistent bias in the predictions. In contrast, reducing the prior mean yields significantly lower predicted lifetimes, but the estimates improve over the course of the experiment and tend to approach the naive single-point predictor $t_f^{\rm pred} = t_{\rm obs}$ at late times. When $t_{\rm obs}$ gets close to the failure time, this actually produces better CRPS-scores (Fig.~\ref{fig:ae_pred_N}b) due to the narrow posterior distribution.

To contrast the effect of prior location with prior width, we additionally consider a less informative prior by doubling the standard deviation while keeping the mean fixed. This modification does not significantly alter the central prediction, but leads to a broader posterior distribution. This indicates that a reasonable ballpark estimate suffices, whereas strongly biased priors can significantly degrade predictive performance.

Finally, we note that prior specification is expected to be material- and measurement-specific. Factors such as the AE sensor characteristics, detection thresholds, and signal processing protocols can influence the effective mapping between observed events and the latent variables. In practice, priors can be elicited from a limited set of preliminary experiments. Importantly, these need not be full creep-to-failure tests: for example monotonic loading experiments may already provide approximate information about e.g. the number of acoustic events.

\end{document}